%% file: main.tex
\documentclass[conference]{IEEEtran}
\IEEEoverridecommandlockouts
\input{package}

\begin{document}
% Abbreviations and self defined commands
\input{HL_new_commands}
\title{High Girth Spatially-Coupled LDPC Codes \\ with Hierarchical Structure\\
% {\footnotesize \textsuperscript{*}Note: Subtitles are not captured in Xplore and
}

\author{%
  \IEEEauthorblockN{Haizheng Li, Sisi Miao, and Laurent Schmalen}
  \IEEEauthorblockA{Karlsruhe Institute of Technology (KIT),
Communications Engineering Lab (CEL),
76187 Karlsruhe, Germany\\
Email: {\{\texttt{haizheng.li@kit.edu\}}}}\vspace{-1em}\thanks{This work received funding from the German Federal Ministry of Education and Research (BMBF) under grant agreement 16KIS2081 (PONGO).}
}

\maketitle

\begin{abstract}
\Ac{qc} \ac{ldpc} codes are a class of LDPC codes with a simple construction facilitating hardware implementation while achieving excellent performance. In this paper, we introduce an algorithm that constructs \ac{qc} \ac{sc} LDPC codes with large girth while keeping the constraint length small. The algorithm offers a ``protograph to basegraph'' construction, focusing on finding small lifting sizes of \ac{qc} codes while avoiding short cycles. This work extends the \ac{hqc} construction for block LDPC codes proposed by Wang et al. to the spatially coupled case. The construction is based on the \ac{crm} derived from the periodic structure of time-invariant SC-LDPC codes. Numerical results show that the proposed algorithm effectively achieves the target girth with a small lifting factor, enabling low-complexity \ac{sc} code construction.
\end{abstract}

\acresetall

\begin{IEEEkeywords}
channel coding, LDPC codes, spatially-coupled codes, high-girth construction, protograph-based code
\end{IEEEkeywords}

%===================================================================================================
\section{Introduction}
\Ac{ldpc} codes~\cite{G62} are an important family of channel codes with outstanding error correction capabilities. \Ac{sc} \ac{ldpc} codes, or \ac{ldpc} convolutional codes~\cite{FZ99}, yield superior performance due to the threshold saturation phenomenon~\cite{LFZ+09}. They are considered promising in future high-throughput communication systems~\cite{SchmalenHandbook20}, which typically require low error floors. Short cycles are widely regarded as detrimental to the decoding performance, often contributing to the error floor under iterative decoding~\cite{R03}. To mitigate this, extensive research has focused on constructing LDPC codes with large girth, both for block codes~\cite{F04, XEA05, WDY13} and \ac{sc} codes~\cite{CS15, NB21}. The progressive edge growth (PEG) algorithm constructs \ac{ldpc} block codes by greedily optimizing the local girth caused by every edge~\cite{XEA05}. The PEG algorithm was extended to \ac{qc} \ac{ldpc} codes in~\cite{LK04}. Girth-$ 6 $ ~\ac{sc} protographs were constructed in~\cite{CS15} by introducing a stringent structural constraints to the protograph. This construction leads to SC-LDPC codes with girth up to 10; however, the resulting coupling widths are rather large. SC-LDPC codes were also considered in~\cite{WDY13}, where the whole \ac{sc} protograph was treated as a block and undergoes block code construction. This rather trivial extension leads to time-variant SC-LDPC and the construction complexity increases exponentially in the number of spatial positions; both properties are not desired in practical SC-LDPC code designs. In~\cite{NB21}, a construction that eliminates short trapping sets was proposed. It was shown that codes with smaller girth may even outperform those with larger girth when the small trapping sets are eliminated. %\blfootnote{This work received funding from the German Federal Ministry of Education and Research (BMBF) under grant agreement 16KIS2081 (PONGO).}

In this work, we propose to construct high-girth \ac{qc} SC-LDPC codes based on their \emph{\ac{crm}} derived from the time-invariant \ac{sc}-\ac{ldpc} code structure, and then adopt the concept of \ac{hqc} \ac{ldpc} codes~\cite{WDY13}. The construction complexity is minimized by systematically exploiting the constraints imposed by the hierarchical circulant and convolutional structures, leading to girth-optimized low-complexity \ac{sc}-\ac{ldpc} codes with small coupling widths and small code lengths. The construction is efficient even for high-rate codes, which typically have large and dense protographs.

%===================================================================================================
\section{Preliminaries}
\label{sec_preliminaries}
An $ [n,k] $ block binary \ac{ldpc} code is defined by the kernel of a sparse \ac{pcm} $ \bH \in \mathbb{F}_2^{m\times n} $, where $ k=n-\text{rank}(\bH) $ is the code dimension and $ n $ is the block length. The \ac{pcm} of a $ (\dv, \dc) $ regular \ac{ldpc} code has $ \dc $ ones in each row and $ \dv $ ones in each column. The code has the rate $ r=k/n $ and the design rate $ r_{\text{d}} = 1-\dv/\dc $. The code can also be represented by its Tanner graph $ \mathcal{G} = \{ \mathcal{N}, \mathcal{E} \} $~\cite{T81}, which is a bipartite graph with a set of \acp{cn} $ \mathcal{C} $ and a set of \acp{vn} $ \mathcal{V} $ with $\mathcal{N} =  \mathcal{V} \cup \mathcal{C} $. An edge $e_{i,j}\in  \mathcal{E}$ exists between the $i$-th \ac{cn} and the $j$-th \ac{vn} if $H_{i,j}\!=\!1$, where $\!H_{i, j}\!$ is the entry of row $\!i\!$ and column $\!j\!$ of $ \bH $.

A \textit{protograph} $ \mathcal{G}_P = \{ \mathcal{N}_P, \mathcal{E}_P \} $ is a small bipartite graph similar to a Tanner graph but allowing repeated edges. The multiplicity of the edges is then denoted as $ P_{i,j} $, which forms the \textit{protomatrix}  $ \bm{P} \in \mathbb{N}^{M \times N}$. A Tanner graph can be constructed by \ac{qc} lifting with a lifting factor $S$, where we replace each entry $ P_{i, j} $ in the protomatrix with a circulant matrix of size $ S\times S $, which has the row weight $ P_{i, j} $, resulting in a QC LDPC code\cite{F04}. In addition, all the circulant matrices of size $ S\times S $ form a ring that is isomorphic to the ring of polynomials \mbox{$ \mathbb{F}_2[x]/\large\langle x^S-1\large\rangle $}. Therefore, we can equivalently represent a QC LDPC code by a polynomial matrix $ \bH(x) $, with the circulant matrices in $ \bH $ replaced by the corresponding polynomials under the isomorphism. A special case are \mbox{type-I} QC LDPC codes where $ \bH(x) $ contains only zeros or monomials $ x^s \in  \mathbb{F}_2[x]/\large\langle x^S-1\large\rangle $, where $ x^s $ corresponds to the \ac{cpm} that has ones on its $ s $-left shifted diagonal under the isomorphism.

An SC-LDPC code with replication factor $L$ and coupling width $w$ can be constructed from an underlying \ac{ldpc} block code with \ac{pcm} $\bH$ using \textit{edge spreading}~\cite{FZ99}. The underlying Tanner graph is replicated $ L $ times and each copy (denoted \emph{batch}) is indexed from $ 1 $ to $ L $. For every batch, the edges inside of the batch are randomly removed and reconnected to its $ w $ following neighboring batches. From the \ac{pcm} perspective, the process is equivalent to decomposing the block code $ \bH $ into component matrices $ \bH_0, \bH_1, \dots, \bH_{w-1} $, and replicate the resulting matrix $ (\bH_0^\mathsf{T}\ \bH_1^\mathsf{T}\ \dots\ \bH_{w-1}^\mathsf{T})^\mathsf{T} $ on the banded diagonal of the SC-LDPC matrix $ \bH_{\text{SC}} $ for $ L $ times.  The decomposition requires that $ \sum_{i=0}^{w-1} \bH_i=\bH $, where the summation is taken over $ \mathbb{N} $. The \ac{pcm} of such an SC-LDPC code is given by
\vspace{-1em}
\begin{equation*}
	\bH_\text{SC} = \underbrace{ \left(
		\footnotesize \begin{matrix}\bH_0 \\ \vdots \\\bH_{w-1} \\ \\ \\  \\ \end{matrix} \hspace{5pt}
		\footnotesize \begin{matrix} \\ \bH_0  \\ \vdots \\\bH_{w-1} \\ \\ \\  \end{matrix} \hspace{8pt}
		\footnotesize \begin{matrix} \\ \\ \ddots \\ \ddots \\ \ddots \\  \\ \end{matrix} \hspace{8pt}
		\footnotesize \begin{matrix} \\ \\ \\ \bH_0  \\ \vdots \\\bH_{w-1} \\ \end{matrix}
		\right).
	}_{L \text{ replications}}
\end{equation*}

To construct \ac{qc} SC-LDPC codes, we may spread the edges of the protograph $ \bP $ to $ \bP_\text{SC} $, and apply \ac{qc} lifting on $ \bP_\text{SC} $ to construct $ \bH_\text{SC} $ or equivalently $ \bH_\text{SC}(x) $.

%===================================================================================================
\section{Cycles in \ac{ldpc} Codes with Structure}
\label{sec_cycles_in_ldpc}
In a Tanner graph $ \mathcal{G}=\{\mathcal{N}, \mathcal{E}\} $, a cycle of length $ \ell $ is a sequence of distinct nodes $ n_1, n_2, \dots, n_\ell $ with $ \ell\ge4 $, such that the edges $ (n_i, n_{i+1}) \in \mathcal{E} $ for $ i=1, 2, \dots, \ell-1 $ and $ (n_\ell, n_1) \in \mathcal{E} $. Or, we can directly identify a cycle as the sequence of edges in $ \mathcal{G} $. A graph has \textit{girth}~$ g $ when it does not contain any cycles of length shorter than~$ g $. In what follows, we highlight how to simplify the cycle identification by exploiting the structure induced by \ac{qc} lifting and edge spreading in SC-LDPC codes.

For block QC LDPC codes, finding a cycle in the Tanner graph can be reduced to finding a cycle in the polynomial matrix $ \bH(x) $. A closed path consisting of $ \ell $ edges $ x^{s_1} \rightarrow x^{s_2} \rightarrow \dots \rightarrow x^{s_\ell} \rightarrow x^{s_1} $ from $ \bH(x) $ form a cycle iff~\cite{F04}
\begin{equation}\label{equ_cycle_condition_Hx}
	\textstyle \sum_{i=1}^\ell (-1)^i s_i \equiv 0 \bmod{S}.
\end{equation}
To find all cycles of length $ \ell $, we need to check all closed paths of length $ \ell $ in $ \bH(x) $ using condition \eqref{equ_cycle_condition_Hx}. This enumerating step is usually done with a depth-first search (DFS) algorithm.

For SC-LDPC codes, the edge spreading procedure never increases the girth~\cite{MDC14}. Furthermore, the convolutional structure of time-invariant SC-LDPC codes leads to a periodic pattern of the cycles. Therefore, for cycle detection in SC-LDPC codes, we don't need to consider the complete Tanner graph consisting of $ L $ replications of subgraphs. Instead, we propose to consider a \textit{cycle relevant submatrix (CRM)} $ \bH_\text{SC}^{[w, \ell]} $ for cycles of lengths $ \ell $. We define the \ac{crm} as the first $ m\left(1+\lfloor \frac{\ell+2}4\rfloor (w-1)\right) $ rows and first $ n\left(1+\lfloor \frac{\ell}4\rfloor (w-1)\right) $ columns of $ \bH_\text{SC} $, with $ m $ and $ n $ being the size of the underlying batch. This follows by considering the largest possible span of a cycle of length $ \ell $, i.e., $ \bH_\text{SC}^{[w, \ell]} $ is guaranteed to contain all the cycle patterns of length $ \ell $. For example, 
\begin{equation*}%\label{equ_cycle_in_scldpc}
    \bH_\text{SC}=\left(\hspace{5pt}
    \footnotesize \begin{matrix}
    \bH_0 & & & & \\
    \bH_1 & \bH_0 & & & \\
    \bH_2 & \bH_1 & \bH_0 & & \\
          & \bH_2 & \bH_1 & \bH_0 & \\
	   &   & \bH_2 & \bH_1 & \bH_0 & \\[-8.5pt]
	   & 	   &   & \bH_2 & \bH_1 & \scriptsize\ddots\\[-8.5pt]
	   & 	   & 	 &   & \bH_2 & \scriptsize\ddots \\[-8.5pt]
	   & & & & & \scriptsize\ddots
	\end{matrix}
	\right)
    \begin{tikzpicture}[overlay, remember picture]
	\node (O) at (-7.6, -2.36) {};
	\draw[KITred, line width=1.5pt] ([xshift=84, yshift=107] O.center) rectangle ([xshift=144, yshift=79] O.center);
	\draw[KITblue, line width=1.5pt] ([xshift=83, yshift=108] O.center) rectangle ([xshift=145, yshift=62] O.center);
	\draw[KITgreen!80, line width=1.5pt] ([xshift=82, yshift=109] O.center) rectangle ([xshift=189, yshift=61] O.center);
	\draw[KITorange, line width=1.5pt] ([xshift=81, yshift=110] O.center) rectangle ([xshift=190, yshift=43] O.center);
	
	\node[KITred] at ([xshift=128, yshift=102] O.center) {\footnotesize $4$-cycle};
	\node[KITblue] at ([xshift=97, yshift=66] O.center) {\footnotesize $6$-cycle};
	\node[KITgreen] at ([xshift=174, yshift=102] O.center) {\footnotesize $8$-cycle};
	\node[KITorange] at ([xshift=97, yshift=48] O.center) {\footnotesize $10$-cycle};
    \end{tikzpicture}
\end{equation*} 
is the \ac{pcm} of an SC-LDPC code with coupling width $ w=3 $. The \acp{crm} for cycles of length $ 4, 6, 8 $ and $ 10 $ are marked with rectangles of different colors. The whole \ac{pcm} $ \bH_\text{SC} $ is free of cycles of length $ \ell $ iff $ \bH_\text{SC}^{[w, \ell]} $ contains no $ \ell $-cycles.

%===================================================================================================
\section{HQC LDPC Codes}
In this section, we briefly review the HQC LDPC code construction~\cite{WDY13}, which enables efficient girth optimization for QC LDPC codes. We only consider HQC LDPC codes obtained by two-stage lifting. It is straightforward to extend the proposed algorithm to the multi-stage lifting case. Then, we also review the algorithm of girth optimization based on the HQC construction.

\subsection{HQC LDPC Code Construction}
The difference between HQC LDPC codes and general \mbox{QC LDPC} codes is that the former possesses a polynomial matrix with a QC structure. For an exemplary QC LDPC code with lifting factor $S = 10$, 
\renewcommand{\arraycolsep}{3pt}
\renewcommand{\arraystretch}{1.2}
\begin{equation}\label{equ_HQC_example_x}
	\hspace{-1.1em}\boldsymbol{H}(x) = 
	\left( 
	\footnotesize \begin{array}{ccc|ccc|c@{\hspace{-5pt}}c@{\hspace{-6pt}}c}
		0 & 0 & x^2 & x^5 & x^7 & 0 & 0 & 0 & 0 \\
		x^2 & 0 & 0 & 0 & x^5 & x^7 & 0 & 0 & 0 \\
		0 & x^2 & 0 & x^7 & 0 & x^5 & 0 & 0 & 0 \\
		\hline
		x^1 & 0 & 0 & 0 & x^8 & 0 & 0 & 0 & x^3+x^8 \\
		0 & x^1 & 0 & 0 & 0 & x^8 & x^3+x^8 & 0 & 0 \\
		0 & 0 & x^1 & x^8 & 0 & 0 & 0 & x^3+x^8 & 0
	\end{array}
	\right)
\end{equation}
is the polynomial matrix of it. The polynomial matrix can be divided into six submatrices of size $ 3\times3 $, which are also circulant.
% in the sense that the polynomial entries on the same shifted diagonal are all identical. 
A new dummy variable $ y $ is now introduced to present this additional structure, and we can write the bivariate polynomial matrix corresponding to~\eqref{equ_HQC_example_x} as
\begin{equation}%\label{equ_HQC_example_xy}
	\boldsymbol{H}(x, y) = \left( 
	\begin{matrix}
		y^1x^2 & \ y^0x^5+y^2x^7\  & 0\\
		y^0x^1 & y^2x^8 & y^1(x^3+x^8)
	\end{matrix}
	\right).
\end{equation}
To distinguish the two levels of lifting, we denote by ``$ y $-lifting'' the lifting from $ \bH(x, y) $ to $ \bH(x) $  with a $y$-lifting factor $ S_y $ such that the circulant submatrices in $\bm{H}(x)$ are of size $S_y \times S_y$. In the above example, $ S_y=3 $. Similarly, the procedure of lifting from $ \bH(x) $ to the QC LDPC \ac{pcm} $ \bH $  is referred to as ``$ x $-lifting'' with ($x$-)lifting factor $ S_x=S=10 $.
% Each monomial in $ \bH(x, y) $ stands for a shifted diagonal in the corresponding $ S_y \times S_y $ submatrix in $ \bH(x) $ consisting of monomials in $ x $, and thus a permutation matrix of size $ S_xS_y \times S_xS_y $ in $ \bH $. Two permutation matrices are either disjoint or identical. Therefore, 
Each monomial in the entries of $ \bH(x, y) $ leads to a permutation matrix of size $ S_xS_y \times S_xS_y $ in $ \bH $. 
The number of terms in a polynomial entry in $ \bH(x, y) $ gives the edge multiplicity in the protograph. The protograph $ \bP $ of the example is \(\bP =  \tiny \begin{psmallmatrix}
	1 & 2 & 0\\
	1 & 1 & 2
\end{psmallmatrix}\).

In the following, we consider only the case where after $ y $-lifting, $ \bH(x) $ contains only monomials, i.e., the terms in one polynomial in $ \bH(x, y) $ should have different $ y $-exponents, leading to type-I QC LDPC codes.

\subsection{Algorithm Description}
\label{sub_sec_algorithm_description}
To find cycles of length $ \ell $ in $ \bH(x, y) $ with lifting factors $ S_x $ and $ S_y $, we need to start with a length-$ \ell $ closed path given by a series of end-to-end connected edges (monomials) $ x^{s_{x,1}}y^{s_{y,1}} \rightarrow x^{s_{x,2}}y^{s_{y,2}} \rightarrow \dots \rightarrow x^{s_{x,\ell}}y^{s_{y,\ell}} \rightarrow x^{s_{x,1}}y^{s_{y,1}} $. Such a closed path forms a cycle iff
\begin{equation*}% \label{equ_cycle_condition_Hxy}
	\left\{
	\begin{aligned}
		\Sigma_x = \textstyle \sum_{i=1}^{\ell} (-1)^i s_{x, i} \equiv 0 \bmod{S_x} \\
		\Sigma_y = \textstyle \sum_{i=1}^{\ell} (-1)^i s_{y, i} \equiv 0 \bmod{S_y}
	\end{aligned}
	\right.	
\end{equation*}
holds. Based on the bivariate polynomial matrix $ \bH(x, y) $, the number of closed paths to be checked for cycle enumeration reduces exponentially in the size of $ \bH(x, y) $ compared to traditional QC LDPC codes, which is key to the high efficiency of the \ac{hqc} construction algorithm.

\begin{figure}
	\centering
	\input{figures/cost_function_table}
	\caption{Cost function tables for optimizing the exemplary HQC LDPC code specified by~\eqref{equ_HQC_exponent_undetermined}}.
	\label{fig_cost_function_table}
\end{figure}

For a given protograph $ \bP $,  lifting factors $ S_x $ and $ S_y $, and target girth $ g $, the goal of the optimization algorithm is to construct an HQC LDPC code belonging to the protograph ensemble with girth~$ g $. Based on $ \bP $, we can already determine the form of $ \bH(x, y) $, with the exponents of the indeterminates $ x $ and $ y $ left unknown. For example, consider again the example \(\boldsymbol{P} =  \tiny \begin{psmallmatrix}
	1 & 2 & 0\\
	1 & 1 & 2
\end{psmallmatrix}\) and write
\begin{equation}\label{equ_HQC_exponent_undetermined}
	\bH(x, y) \!= \!\left(\!
	\begin{array}{cc@{\hspace{-6pt}}c}
		x^{a_1} y^{b_1} & x^{a_2} y^{b_2} + x^{a_3} y^{b_3} & 0 \\
		x^{a_4} y^{b_4} & x^{a_5} y^{b_5} & x^{a_6} y^{b_6} + x^{a_7} y^{b_7}
	\end{array}\!
	\right),
\end{equation}
where $a_1, \ldots, a_7,b_1,\ldots, b_7$ are exponents to be optimized. The optimization problem can be formulated as:
\begin{align}
	(a_1, \dots, a_7, b_1, \dots, b_7) =&  \arg \min_{\substack{(a_1, \ldots, a_7)\\ (b_1, \ldots, b_7)}} \quad  \text{girth}(\bH(x, y)) \notag\\
	\text{subject to} \quad &	0 \le a_1, \dots, a_7 \le S_x-1, \notag\\
	& 0 \le b_1, \dots, b_7 \le S_y-1, \notag\\
	& b_2\ne b_3, b_6\ne b_7. \notag
\end{align}

The \ac{hqc} construction algorithm initializes the unknown exponents randomly and optimizes them with a 
greedy strategy. As depicted in Fig.~\ref{fig_cost_function_table}, we maintain two tables of cost function values for both $x$ and $y$ exponents, respectively. The entries of the tables are calculated as follows. For each unknown exponent, e.g., $ a_1 $, and each of its possible values $ s \in \{0, 1, \dots, S_x-1\} $, we fix all other exponents $ a_2, \dots, a_7, b_1, \dots, b_7 $ to their current values, and we count the number of short cycles up to length $ g-2 $ when we set $ a_1 = s $. Let $ N_4, N_6, \dots, N_{g-2} $ be the number of cycles of length $ 4, 6, \dots, g-2 $ caused by setting $ a_1 = s $. The entry $C_{a_1, s}$ is calculated as a weighted sum $C_{a_1, s} =\textstyle \sum_{i=4}^{g-2} w_i N_i$, where the weights $ w_i $ are to be optimized. We empirically choose the weights $ w_i $ as powers of $ 5 $, i.e., $ w_4 = 5^3, w_6 = 5^2, w_8 = 5, w_{10} = 1 $. We can then fill up the table of the exponents of $ x $ using this approach. For the table of the exponents of $ y $, the same approach can be applied except for the additional requirement that the exponents of the $ y $-terms in the same polynomial should be different. Values that cause repeated exponents of $ y $ have cost of infinity.

\begin{algorithm}[t]
	\caption{\ac{hqc} Construction}
	\begin{algorithmic}[1]
		\State \textbf{Input:} Protograph $ \bP $, lifting factors $ S_x $ and $ S_y $, girth $g$
		\Procedure{HQC\_CONSTR}{$ \bP,\ S_x,\ S_y,\ g $}
		\State Transform $ \bP $ to bivariate polynomials $ \bH(x, y) $
		\State Initialize undetermined exponents randomly
		\While{1}
		\For{$ \ell = 4, 6, \dots, g-2 $}
		\State \textbf{DFS} traverse all closed paths of length $ \ell $
		\State \hspace{1em} Calculate $ \Sigma_x $ and $ \Sigma_y $ of the path
		\State \hspace{1em} Update the cost function tables
		\State \textbf{end DFS}
		\EndFor
		\State Calculate the cost reduction tables
            \If {All entries in both tables nonnegative}
            \State \textbf{break}
            \EndIf
		\State Update the exponent according to Sec.~\ref{sub_sec_algorithm_description}
		\EndWhile
		\State \textbf{return} $ \bH(x, y) $
		\EndProcedure
	\end{algorithmic}
	\label{alg_hqc_construction_block}
\end{algorithm}

After computing the initial state of the two cost function tables, we traverse the table row-wise to find the next exponent that leads to the largest cost reduction. For instance, if the current value  of the exponent $ a_1 $ is $ s_0 $, we subtract $ C_{a_1, s_0} $ from the corresponding row of $ a_1 $ in the cost function table of $ x $. We carry out a similar subtraction for each row of both tables and acquire the cost reduction tables. We find the minimum value among both cost reduction tables and change only \emph{one} exponent to the value corresponding to the minimum. After the change, we update both tables and repeat the process until there are no more negative values in both cost reduction tables. The algorithm is summarized in Alg.~\ref{alg_hqc_construction_block}.

%===================================================================================================
\section{\ac{hqc} for SC-LDPC Construction}
In this section, we introduce two algorithms for constructing \ac{hqc} SC-LDPC codes from given SC protographs.

\subsection{Block Construction and Spread}
The first algorithm exploits the fact that spreading the edges of a block code does not introduce additional cycles\cite{MDC14}. Hence, we can ``despread'' the SC protograph $ \bP_\text{SC} $ back to a block protograph $ \bP $, apply Alg.~\ref{alg_hqc_construction_block} to it, and  spread the resulting block code back to an SC code, which has automatically at least the same girth.
% As long as the construction of the block code $ \bH(x, y) $ fulfills the girth condition, the final spatially coupled code $ \bH_\text{SC}(x, y) $ is also guaranteed to have at least as high girth.
This method is straightforward. However,  when adding the submatrices back to a block protograph, we lose the information about the spreading process, which may itself remove some cycles that originally existed in the block code. Nevertheless, we can make use of this property by constructing $ \bH(x, y) $ with lower girth $ g'=g-2 $ and checking the girth of $ \bH_\text{SC}(x, y) $ after spreading. The success of the construction is probabilistic and we need to repeat the process until the target girth is achieved or some other stopping criterion is met, indicating a failure in construction. The construction algorithm is described in Alg.~\ref{alg_hqc_sc_construction_1}. In fact, none of the algorithms in this paper guarantees a successful construction for an arbitrary protograph, target girth and lifting factors. We need to ``try and check'' for all of the construction algorithms. For simplicity, we leave out this ``try and check'' process in the algorithm descriptions.

\begin{algorithm}[t]
	\caption{Block Construction and Spread}
	\begin{algorithmic}[1]
		\State \textbf{Input:} SC protograph $ \bP_\text{SC} $, coupling width $w$, lifting factors $ S_x $ and $ S_y $, target girth $g$
		\Procedure{HQC\_SC\_CONSTR}{$ \bP_\text{SC}, w, S_x, S_y, g $}
		\State Form the block protograph $ \bP $ by de-spreading
		\State $ \bH(x, y) $=\ac{hqc}\_CONSTR($ \bP, S_x, S_y, g-2) $
		\State Spread $ \bH(x, y) $ to $ \bH_\text{SC}(x, y) $
		\State \textbf{return} $ \bH_\text{SC}(x, y) $
		\EndProcedure
	\end{algorithmic}
	\label{alg_hqc_sc_construction_1}
\end{algorithm}

\subsection{Construction Based on the \ac{crm}}
In Sec.~\ref{sec_cycles_in_ldpc}, we reduced the problem of finding cycles in a QC SC-LDPC code $ \bH_\text{SC} $ to finding cycles only in the bivariate polynomial \ac{crm} $ \bH^{[w, \ell]}_\text{SC}(x, y) $. For the construction of a \mbox{girth-$ g $} code, we need to eliminate all cycles of length up to \mbox{$ g-2 $}, and thus considering $ \bH^{[w, g-2]}_\text{SC}(x, y) $. Basically, the matrix $ \bH^{[w, g-2]}_\text{SC}(x, y) $ fits already into the framework of the \ac{hqc} construction algorithm; however, the undetermined exponents may be replicated due to the replication of the submatrices.

As an example, we take another protograph \( \bP=(2 \quad 3) \) and apply edge spreading with \( w=3 \), yielding \( \bP_0=(1\quad 1) ,  \bP_1=(0\quad 1) \) and \( \bP_2=(1\quad 1)\). The bivariate polynomial matrix $ \bH_\text{SC}(x, y) $ with undetermined exponents is spread correspondingly to \mbox{$ \bH_0(x, y) = (x^{a_1}y^{b_1} \quad x^{a_2}y^{b_2}) $}, \mbox{$ \bH_1(x, y) = (0 \quad x^{a_3}y^{b_3}) $} and \mbox{$ \bH_2(x, y) = (x^{a_4}y^{b_4} \quad x^{a_5}y^{b_5}) $}. The 6-\ac{crm} $ \bH^{[3,6]}_\text{SC}(x, y) $ is given as

\vspace{-1em}
{
\begingroup
	\renewcommand{\arraycolsep}{2.5pt}
	\renewcommand{\arraystretch}{1.2}
        \footnotesize
	\begin{equation*}%\label{equ_HQC_SC_example}
		\bH_\text{SC}^{[3, 6]}(x, y)\! =\!\! \left(
		\begin{array}{cc|cc|cc}
			x^{a_1}y^{b_1} & x^{a_2}y^{b_2} &  &  &  & \\
			0      & x^{a_3}y^{b_3} & x^{a_1}y^{b_1} & x^{a_2}y^{b_2} &  & \\
			x^{a_4}y^{b_4} & x^{a_5}y^{b_5} & 0      & x^{a_3}y^{b_3} & x^{a_1}y^{b_1} & x^{a_2}y^{b_2} \\
			&        & x^{a_4}y^{b_4} & x^{a_5}y^{b_5} & 0      & x^{a_3}y^{b_3} \\
			&        &        &        & x^{a_4}y^{b_4} & x^{a_5}y^{b_5}
		\end{array}
		\right).
	\begin{tikzpicture}[overlay, remember picture]
		\node (O) at (-8.72, -2.16) {};
		\draw[KITred] ([xshift=98, yshift=76] O.center) rectangle ([xshift=130, yshift=65] O.center);
		\draw[KITred] ([xshift=161, yshift=54] O.center) rectangle ([xshift=130, yshift=65] O.center);
	\end{tikzpicture}
    \end{equation*}
    \endgroup
}
Consider the closed path of length $ 6 $ $ x^{a_3}y^{b_3} \rightarrow x^{a_5}y^{b_5} \rightarrow x^{a_3}y^{b_3} \rightarrow x^{a_5}y^{b_5} \rightarrow x^{a_4}y^{b_4} \rightarrow x^{a_1}y^{b_1} $ highlighted in the matrix. By checking the cycle condition, we have
\begin{align*}
	& \Sigma_x = a_3-a_5+a_3-a_5+a_4-a_1 = 2a_3-2a_5+a_4-a_1 \\
	& \Sigma_y = b_3-b_5+b_3-b_5+b_4-b_1 = 2b_3-2b_5+b_4-b_1.
\end{align*}
In contrary to the block code \ac{hqc} construction, we find that the undetermined exponents $ a_3, a_5, b_3, b_5 $ all appear twice. We define the multiplicity $ \kappa $ of an exponent to be its final coefficient in the expression of $ \Sigma_x $ or $ \Sigma_y $. For the given example, the multiplicities of the exponents $ a_3, a_5, a_4, a_1 $ are $ 2, -2, 1,$ and $ -1 $, respectively. In general, for an exponent of current value $ s_0 $ of the indeterminate $ x $ with multiplicity $ \kappa $, which has current path exponent sum $ \Sigma_x $, the problematic new value $ s' $ that would cause cycles can be computed by solving
\begin{equation*}%\label{equ_problematic_s}
	\kappa(s_0-s') \equiv \Sigma_x \bmod{S_x}
\end{equation*}
with respect to $ s' $, and update the entries of the cost function table corresponding to the problematic values of $ s' $. The same process holds for the exponents of indeterminate $ y $.

Another difference from the block code \ac{hqc} construction is that we don't have to traverse all the closed paths of relevant length, but only those which involve the edges in the first replication of $ (\bH_0^{\mathsf{T}}, \bH_1^{\mathsf{T}}, \dots, \bH_{w-1}^{\mathsf{T}})^{\mathsf{T}} $, i.e. the left most block column of submatrices in $ \bH^{[w, g-2]}_\text{SC}(x, y) $. Thus, the overall complexity of the construction algorithm is reduced. We denote the HQC construction with these two modifications the modified HQC construction (summarized  in Alg.~\ref{alg_hqc_sc_construction_2}). 

\begin{algorithm}[t]
	\caption{HQC Construction Based on the \ac{crm}}
	\begin{algorithmic}[1]
		\State \textbf{Input:} SC protograph $ \bP_\text{SC} $, coupling width $w$, lifting factors $ S_x $ and $ S_y $, target girth $g$
		\Procedure{HQC\_SC\_CONSTR}{$ \bP_\text{SC}, w, S_x, S_y, g $}
		\State Slice \ac{crm} $ \bP^{[w, g-2]}_\text{SC} $ from $ \bP_\text{SC} $
		\State Apply the modified HQC\_CONSTR algorithm:
            \Statex \hspace{11pt} $ \bH(x, y) $=\ac{hqc}\_CONSTR\_MOD($ \bP^{[w, g-2]}_\text{SC}, S_x, S_y, g) $
		\State Replicate $ \bH(x, y) $ to $ \bH_\text{SC}(x, y) $
		\State \textbf{return} $ \bH_\text{SC}(x, y) $
		\EndProcedure
	\end{algorithmic}
	\label{alg_hqc_sc_construction_2}
\end{algorithm}

%===================================================================================================
\section{Numerical Results}
We present numerical results demonstrating the effectiveness of the construction algorithms under different setups. We measure the effectiveness by the minimum lifting size $ S_\text{min} $ required to achieve a specified girth for a given protograph since it is generally easier to meet the girth requirements with a larger lifting factor.
% thus achieving a smaller $ S_\text{min} $ indicates a more efficient construction.

\begin{figure}
	\input{figures/S_min_dv3_g10_SC}
	\vspace{-15pt}
	\caption{Minimum $x$-lifting sizes $ S_\text{min} $ found for different girth-10 regular QC SC-LDPC ensembles with $ w=2 $ and $ S_y=1 $}
	\label{fig_S_min_dv3_g10_SC}
\end{figure}
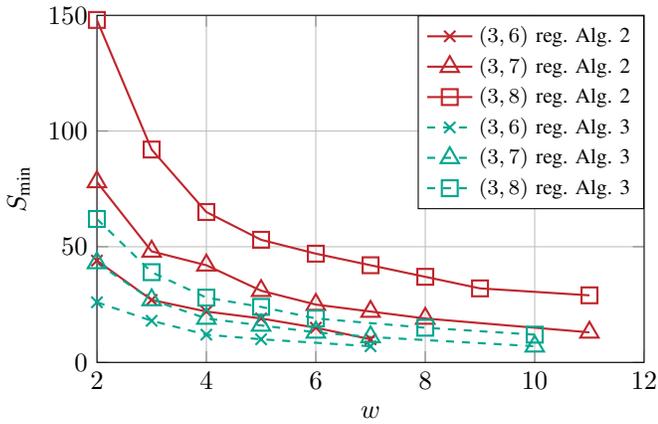

First, we consider QC SC-LDPC codes with a regular $ (\dv=3, \dc) $ underlying block code. The protograph of this underlying block code  has an all-ones matrix of size $ \dv\times \dv $. We set $ \dc $ to $ 6, 7, 8 $, respectively, and sweep over different coupling widths $ w \in \{2, 3, \dots, 12\} $. We then try to construct girth-$ g=10 $ QC SC-LDPC codes with Alg.~\ref{alg_hqc_sc_construction_1} and Alg.~\ref{alg_hqc_sc_construction_2}. We set $ S_y = 1 $ and the minimum lifting size $ S_\text{min}=S_x $ reduces with increasing $ w $. The results are shown in Fig~\ref{fig_S_min_dv3_g10_SC}. %Since the cycles in the underlying code are more likely to be eliminated by spreading when $w$ is large, $ S_\text{min} $ decreases with increasing $w$. 
Alg.~\ref{alg_hqc_sc_construction_2} can achieve lifting factors approximately half as large as those of Alg.~\ref{alg_hqc_sc_construction_1}. The difference becomes even larger for larger $ \dc $. We conclude that Alg.~\ref{alg_hqc_sc_construction_2} is more effective than Alg.~\ref{alg_hqc_sc_construction_1} for constructing high-girth SC-LDPC codes with various lifting factors. % Similar results for $ \dv=4 $ and various $ \dc $ ensembles are presented in Fig.~\ref{fig_S_min_dv4_g10_SC}.

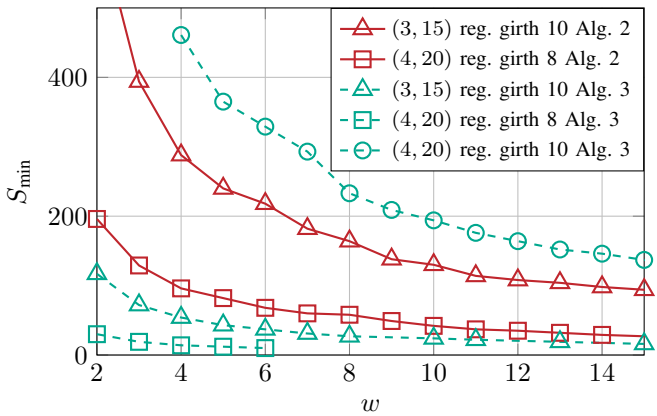
\begin{figure}
	\input{figures/S_min_dv4_dc20_SC}
        \vspace{-15pt}
	\caption{Minimum $x$-lifting sizes $ S_\text{min} $ found for high rate $ r_\text{d}=0.8 $ regular QC SC-LDPC ensembles with $ w=2 $ and $ S_y=4 $}
	\label{fig_S_min_dv4_dc20_SC}
\end{figure}

In many applications, e.g., high-throughput optical communication systems, high-rate codes are required. However, high-rate codes often result in large, dense protographs, complicating code construction. In Fig.~\ref{fig_S_min_dv4_dc20_SC}, we apply both algorithms to construct $ (3, 15) $ and $ (4, 20) $ regular QC SC-LDPC codes with design rate $ r_\text{d}=4/5$, $ y $-lifting size $ S_y=4 $,  target girth $g\in\{8,10\}$ and protographs consisting of a $\dv\times\dc$ all-one matrix. Alg.~\ref{alg_hqc_sc_construction_2} still largely outperforms Alg.~\ref{alg_hqc_sc_construction_1}, and for the $ g=10 $ case, no meaningful results can be obtained by Alg.~\ref{alg_hqc_sc_construction_1}. We highlight that our proposed algorithm yields an SC-LDPC code of $ S_\text{min} = 461 $, yielding a batch size of $ S_\text{min}\cdot S_y\cdot \dc = 36880 $ and $ w=4 $, which is considered practical.

%\begin{figure}
%	\input{figures/S_min_dv4_g10_SC}
%	\caption{Minimum lifting sizes $S_\text{min}$ found for different girth 10 SC-LDPC ensembles}
%	\label{fig_S_min_dv4_g10_SC}
%\end{figure}

%\begin{figure}
%	\input{figures/S_min_dv34_g10_SC}
%	\caption{Minimum lifting sizes $S_\text{min}$ found for different girth 10 SC-LDPC ensembles}
%	\label{fig_S_min_dv34_g10_SC}
%\end{figure}

\begin{figure}
	\input{figures/comparison_dv3_dv6_g6_g10}
    \vspace{-10pt}
    \caption{BER comparison of $(3, 6)$ regular SC-LDPC codes with girth 6 and 10, coupling width $ w=2 $, lifting size $ S=30 $}
	\label{fig_comparison_dv3_dv6_g6_g10}
\end{figure}
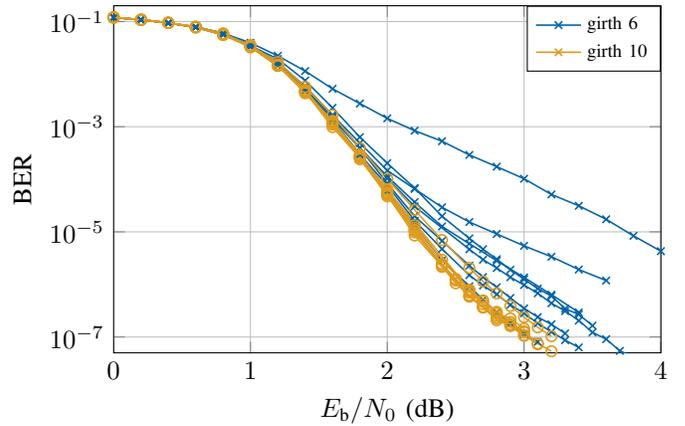

In Fig.~\ref{fig_comparison_dv3_dv6_g6_g10}, we compare multiple realizations of \mbox{$ (3, 6) $}  regular QC SC-LDPC codes of coupling width $ w=2 $ with girth $ 6 $ and $ 10 $, respectively, constructed with lifting factors $ S_x=30 $ and $ S_y=1 $ from the same protograph (all-one matrix). The codes of girth $ 10 $ perform far better on average than the codes of girth $ 6 $, which validates our motivation for constructing codes with higher girth. % As a benchmark we take the code from~\cite{NB21} in Table III, which is also $(3, 6)$ regular. We choose the window size so that it has the same number of bits as the simulation of our proposed method.

%===================================================================================================
\section{Conclusion}
In this work, we proposed two new algorithms for constructing SC-LDPC codes based on the \ac{hqc} framework. The first algorithm extends existing block \ac{hqc} \ac{ldpc} constructions in a straightforward manner, while the second leverages the intrinsic convolutional structure of SC-LDPC codes. Numerical results demonstrate that the latter approach yields more efficient codes, offering a practical and scalable method for designing high-girth QC SC-LDPC codes.


%\bibliographystyle{IEEEtran}
%\bibliography{references}

\end{document}

%% file: package.tex
\usepackage{textcomp}
\usepackage{cite}
\usepackage{amsmath,amssymb,amsfonts}
\usepackage{algorithm}
\usepackage{algorithmicx}
\usepackage{algpseudocode}
\algrenewcommand{\algorithmiccomment}[1]{\hskip3em// #1}

\usepackage{graphicx}
\usepackage{textcomp}
\usepackage{xcolor}
\usepackage{KITcolors}
\usepackage{todonotes}
\usepackage{acronym}
\usepackage{bm}
\usepackage{tikz, pgfplots}
\usetikzlibrary{arrows,calc,fit,matrix,positioning,shapes,shadows,trees,mindmap,tikzmark,arrows.meta,angles,quotes,babel,spy}
\usepackage{float} % Add this in the preamble

\tikzstyle{every picture}+=[remember picture]
\pgfplotsset{width=10cm,compat=1.9}
\usetikzlibrary{decorations.pathreplacing}

\def\BibTeX{{\rm B\kern-.05em{\sc i\kern-.025em b}\kern-.08em
    T\kern-.1667em\lower.7ex\hbox{E}\kern-.125emX}}

    \newenvironment{psmallmatrix}
  {\left(\begin{smallmatrix}}
  {\end{smallmatrix}\right)}

%% file: HL_new_commands.tex
% Define abbreviations
% \makeglossaries

\acrodef{ldpc}[LDPC]{low-density parity-check}
\acrodef{sc}[SC]{spatially-coupled}
\acrodef{qc}[QC]{quasi-cyclic}
\acrodef{hqc}[HQC]{hierarchical quasi-cyclic}

\acrodef{vn}[VN]{variable node}
\acrodefplural{vn}[VNs]{variable nodes}
\acrodef{cn}[CN]{check node}
\acrodefplural{cn}[CNs]{check nodes}

\acrodef{bsc}[BSC]{binary symmetric channel}
\acrodef{bec}[BEC]{binary erasure channel}
\acrodef{awgn}[AWGN]{additive white Gaussian noise}
\acrodef{bi-awgn}[BI-AWGN]{Binary-Input Additive White Gaussian Noise}

\acrodef{ber}[BER]{bit error rate}
\acrodef{fer}[FER]{frame error rate}
\acrodef{snr}[SNR]{signal-to-noise ratio}

\acrodef{pcm}[PCM]{parity-check matrix}
\acrodefplural{pcm}[PCMs]{parity check matrices}
\acrodef{cpm}[CPM]{cyclic permutation matrix}
\acrodefplural{cpm}[CPMs]{cyclic permutation matrices}

\acrodef{crm}[CRM]{cycle relevant matrix}
\acrodefplural{crm}[CRMs]{cycle relevant matrices}

\newcommand{\bB}{\boldsymbol{B}}
\newcommand{\bH}{\boldsymbol{H}}
\newcommand{\bI}{\boldsymbol{I}}
\newcommand{\bP}{\boldsymbol{P}}
\newcommand{\bQ}{\boldsymbol{Q}}
\newcommand{\dv}{d_\text{v}}
\newcommand{\dc}{d_\text{c}}
\newcommand{\svdots}{\raisebox{3pt}{$\scalebox{.75}{\vdots}$}}
\newcommand{\sddots}{\raisebox{3pt}{$\scalebox{.75}{$\ddots$}$}} 

%% file: figures/cost_function_table.tex
\begin{tikzpicture}[scale=0.9]
	\def\step{21pt}
	\node (a1) at (0, 5) {$ a_1 $};
	\node (a2) at ([yshift=-\step] a1.center) {$ a_2 $};
	\node (a3) at ([yshift=-\step] a2.center) {$ a_3 $};
	\node (a4) at ([yshift=-\step] a3.center) {$ a_4 $};
	\node (a5) at ([yshift=-\step] a4.center) {$ a_5 $};
	\node (a6) at ([yshift=-\step] a5.center) {$ a_6 $};
	\node (a7) at ([yshift=-\step] a6.center) {$ a_7 $};
	
	\node at ([xshift=1*\step, yshift=0.9*\step] a1.center) {$ 0 $};
	\node at ([xshift=2*\step, yshift=0.9*\step] a1.center) {$ 1 $};
	\node[font=\footnotesize] at ([xshift=5.2*\step, yshift=0.9*\step] a1.center) {$ S_x\!-\!1 $};
	
	\fill[black!10] ([xshift=0.5*\step, yshift=0.5*\step] a1.center) rectangle ([xshift=2.5*\step, yshift=-6.5*\step] a1.center);
	\fill[black!10] ([xshift=4.7*\step, yshift=0.5*\step] a1.center) rectangle ([xshift=5.7*\step, yshift=-6.5*\step] a1.center);
	
	\foreach \i in {1, 2, ..., 7}
		\node at ([xshift=3.67*\step, yshift=1*\step-\step*\i] a1.center) {$ \dots $};
	
	\node[font=\tiny] at ([xshift=1*\step, yshift=-0.1*\step] a1.center) {$ C_{a_1,0} $};
	\node[font=\tiny] at ([xshift=2*\step, yshift=-0.1*\step] a1.center) {$ C_{a_1,1} $};
	\node[font=\tiny] at ([xshift=1*\step, yshift=-0.1*\step] a2.center) {$ C_{a_2,0} $};
	\node[font=\tiny] at ([xshift=2*\step, yshift=0.1*\step] a2.center) {$ \ddots $};
	\node[font=\tiny] at ([xshift=1*\step, yshift=0.1*\step] a3.center) {$ \vdots $};
	\node[font=\tiny] at ([xshift=2*\step, yshift=0.1*\step] a3.center) {$ \ddots $};
	
	\foreach \i in {1, 2, ..., 8}{
		\draw ([xshift=0.5*\step, yshift=1.5*\step-\step*\i] a1.center) -- ([xshift=3.0*\step, yshift=1.5*\step-\step*\i] a1.center);
		\draw ([xshift=4.2*\step, yshift=1.5*\step-\step*\i] a1.center) -- ([xshift=5.7*\step, yshift=1.5*\step-\step*\i] a1.center);
	}
	\foreach \i in {1, 2, 3}
	\draw ([xshift=-0.5*\step+\step*\i, yshift=1.5*\step-\step] a1.center) -- ([xshift=-0.5*\step+\step*\i, yshift=1.5*\step-\step*8] a1.center);
	
	\foreach \i in {1, 2}
	\draw ([xshift=3.7*\step+\step*\i, yshift=1.5*\step-\step] a1.center) -- ([xshift=3.7*\step+\step*\i, yshift=1.5*\step-\step*8] a1.center);

%%%%%%%%%%%%%%%%%%%%%%%%%%%%%%%%%%%%%%%%%%%%%%%%%%%%%%%%%%%%%%%%%%%%%%%%
	\node (b1) at (5.2, 5) {$ b_1 $};
	\node (b2) at ([yshift=-\step] b1.center) {$ b_2 $};
	\node (b3) at ([yshift=-\step] b2.center) {$ b_3 $};
	\node (b4) at ([yshift=-\step] b3.center) {$ b_4 $};
	\node (b5) at ([yshift=-\step] b4.center) {$ b_5 $};
	\node (b6) at ([yshift=-\step] b5.center) {$ b_6 $};
	\node (b7) at ([yshift=-\step] b6.center) {$ b_7 $};
	
	\node at ([xshift=1*\step, yshift=0.9*\step] b1.center) {$ 0 $};
	\node at ([xshift=2*\step, yshift=0.9*\step] b1.center) {$ 1 $};
	\node[font=\footnotesize] at ([xshift=5.2*\step, yshift=0.9*\step] b1.center) {$ S_y\!-\!1 $};
	
	\foreach \i in {1, 2, ..., 7}
	\node at ([xshift=3.67*\step, yshift=1*\step-\step*\i] b1.center) {$ \dots $};
	
	\fill[black!10] ([xshift=0.5*\step, yshift=0.5*\step] b1.center) rectangle ([xshift=2.5*\step, yshift=-6.5*\step] b1.center);
	\fill[black!10] ([xshift=4.7*\step, yshift=0.5*\step] b1.center) rectangle ([xshift=5.7*\step, yshift=-6.5*\step] b1.center);
	
	\node[font=\tiny] at ([xshift=1*\step, yshift=-0.1*\step] b1.center) {$ C_{b_1,0} $};
	\node[font=\tiny] at ([xshift=2*\step, yshift=-0.1*\step] b1.center) {$ C_{b_1,1} $};
	\node[font=\tiny] at ([xshift=1*\step, yshift=-0.1*\step] b2.center) {$ C_{b_2,0} $};
	\node[font=\tiny] at ([xshift=2*\step, yshift=0.1*\step] b2.center) {$ \ddots $};
	\node[font=\tiny] at ([xshift=1*\step, yshift=0.1*\step] b3.center) {$ \vdots $};
	\node[font=\tiny] at ([xshift=2*\step, yshift=0.1*\step] b3.center) {$ \ddots $};
	
	\foreach \i in {1, 2, ..., 8}{
		\draw ([xshift=0.5*\step, yshift=1.5*\step-\step*\i] b1.center) -- ([xshift=3.0*\step, yshift=1.5*\step-\step*\i] b1.center);
		\draw ([xshift=4.2*\step, yshift=1.5*\step-\step*\i] b1.center) -- ([xshift=5.7*\step, yshift=1.5*\step-\step*\i] b1.center);
	}
	\foreach \i in {1, 2, 3}
		\draw ([xshift=-0.5*\step+\step*\i, yshift=1.5*\step-\step] b1.center) -- ([xshift=-0.5*\step+\step*\i, yshift=1.5*\step-\step*8] b1.center);
		
	\foreach \i in {1, 2}
		\draw ([xshift=3.7*\step+\step*\i, yshift=1.5*\step-\step] b1.center) -- ([xshift=3.7*\step+\step*\i, yshift=1.5*\step-\step*8] b1.center);
\end{tikzpicture}

%% file: figures/S_min_dv3_g10_SC.tex
\begin{tikzpicture}[scale=1]% [spy using outlines={circle, magnification=6, connect spies}]
	\begin{axis}[
		width=\columnwidth,
		height=0.7\columnwidth,
		xlabel={$w$},    
		ylabel={$ S_\text{min} $},
		 xmin=2, xmax=12,
		 ymin=0, ymax=150,
		ymajorgrids=true,
		xmajorgrids=true,
		legend cell align={left},
		legend style={at={(1, 1)},anchor=north east,font=\footnotesize},
		]
		%---------------------------------------------
		\addplot[
		color=KITred,
		thick,
		mark=x,
		mark size=3pt,
		%forget plot,
		]
		coordinates{
			(2, 44)
                (3, 27)
                (4, 22)
                (5, 19)
                (6, 15)
                (7, 10)
		};
		\addlegendentry{$(3, 6)$ reg. Alg.~\ref{alg_hqc_sc_construction_1}}
            %---------------------------------------------
		\addplot[
		color=KITred,
		thick,
		mark=triangle,
		mark size=4pt,
		%forget plot,
		]
		coordinates{
			(2, 78)
                (3, 48)
                (4, 42)
                (5, 31)
                (6, 25)
                (7, 22)
                (8, 19)
                (11, 13)
		};
		\addlegendentry{$(3, 7)$ reg. Alg.~\ref{alg_hqc_sc_construction_1}}
            %---------------------------------------------
		\addplot[
		color=KITred,
		thick,
		mark=square,
		mark size=3pt,
		%forget plot,
		]
		coordinates{
			(2, 148)
                (3, 92)
                (4, 65)
                (5, 53)
                (6, 47)
                (7, 42)
                (8, 37)
                (9, 32)
                (11, 29)
		};
		\addlegendentry{$(3, 8)$ reg. Alg.~\ref{alg_hqc_sc_construction_1}}
%%%%%%%%%%%%%%%%%%%%%%%%%%%%%%%%%%%%%%%%%%%%%%%%%%%%%%%%%%%%%%%%%%%%%%%%%%%%%%%%%%%%%%%
		%---------------------------------------------
		\addplot[
		color=KITgreen,
		thick,
            dashed,
		mark=x,
		mark size=3pt,
            mark options={solid}
		%forget plot,
		]
		coordinates{
			(2, 26)
                (3, 18)
                (4, 12)
                (5, 10)
                (7, 7)
		};
		\addlegendentry{$(3, 6)$ reg. Alg.~\ref{alg_hqc_sc_construction_2}}
            %---------------------------------------------
		\addplot[
		color=KITgreen,
		thick,
            dashed,
		mark=triangle,
		mark size=4pt,
            mark options={solid}
		%forget plot,
		]
		coordinates{
			(2, 43)
                (3, 27)
                (4, 19)
                (5, 16)
                (6, 13)
                (7, 11)
                (10, 7)
		};
		\addlegendentry{$(3, 7)$ reg. Alg.~\ref{alg_hqc_sc_construction_2}}
            %---------------------------------------------
		\addplot[
		color=KITgreen,
		thick,
            dashed,
		mark=square,
		mark size=3pt,
            mark options={solid}
		%forget plot,
		]
		coordinates{
			(2, 62)
                (3, 39)
                (4, 28)
                (5, 24)
                (6, 19)
                (8, 15)
                (10, 12)
		};
		\addlegendentry{$(3, 8)$ reg. Alg.~\ref{alg_hqc_sc_construction_2}}
	\end{axis}
%	\spy [blue, size=2.5cm] on (axis cs:2.8, 1e-3) in node[fill=white] at (axis cs:2.4, 1e-4);
\end{tikzpicture}

%% file: figures/S_min_dv4_dc20_SC.tex
\begin{tikzpicture}[scale=1]% [spy using outlines={circle, magnification=6, connect spies}]
	\begin{axis}[
		width=\columnwidth,
		height=0.7\columnwidth,
		xlabel={$w$},    
		ylabel={$ S_\text{min} $},
		xmin=2, xmax=15,
		ymin=0, ymax=500,
		ymajorgrids=true,
		xmajorgrids=true,
		legend cell align={left},
		legend style={at={(1, 1)},anchor=north east,font=\footnotesize},
		]
            %---------------------------------------------
		\addplot[
		color=KITred,
		thick,
		mark=x,
		mark size=3pt,
		forget plot,
            opacity=0,
		]
		coordinates{
			(2, 20)
			(3, 14)
			(4, 10)
		};
		% \addlegendentry{$(3, 15)$ reg. girth 8 Alg.~\ref{alg_hqc_sc_construction_1}}
            %---------------------------------------------
		\addplot[
		color=KITred,
		thick,
		mark=triangle,
		mark size=4pt,
		%forget plot,
		]
		coordinates{
			(2, 630)
			(3, 394)
			(4, 288)
                (5, 240)
                (6, 218)
                (7, 182)
                (8, 164)
                (9, 138)
                (10, 130)
                (11, 114)
                (12, 108)
                (13, 104)
                (14, 98)
                (15, 94)
                (16, 86)
                (17, 82)
                (18, 74)
		};
		\addlegendentry{$(3, 15)$ reg. girth 10 Alg.~\ref{alg_hqc_sc_construction_1}}
		%---------------------------------------------
		\addplot[
		color=KITred,
		thick,
		mark=square,
		mark size=3pt,
		%forget plot,
		]
		coordinates{
			(2, 196)
			(3, 129)
			(4, 96)
			(5, 82)
			(6, 68)
			(7, 60)
			(8, 58)
			(9, 49)
			(10, 42)
			(11, 37)
			(12, 35)
			(13, 32)
			(14, 29)
			(16, 25)
			(17, 23)
			(18, 21)
			(23, 15)
		};
		\addlegendentry{$(4, 20)$ reg. girth 8 Alg.~\ref{alg_hqc_sc_construction_1}}
        %%%%%%%%%%%%%%%%%%%%%%%%%%%%%%%%%%%%%%%%%%%%%%%%%%%%%%%%%%%%%%%%%%%%%%%%%%%%%%%%%%%%%%%
            %---------------------------------------------
		\addplot[
		color=KITgreen,
		thick,
            dashed,
		mark=x,
		mark size=3pt,
		forget plot,
            opacity=0,
		]
		coordinates{
			(2, 7)
			(3, 5)
		};
		% \addlegendentry{$(3, 15)$ reg. girth 8 Alg.~\ref{alg_hqc_sc_construction_2}}
            %---------------------------------------------
		\addplot[
		color=KITgreen,
		thick,
            dashed,
		mark=triangle,
		mark size=4pt,
            mark options={solid}
		%forget plot,
		]
		coordinates{
			(2, 117)
			(3, 72)
			(4, 54)
                (5, 43)
                (6, 37)
                (7, 31)
                (8, 27)
                (10, 24)
                (11, 22)
                (13, 19)
                (15, 16)
		};
		\addlegendentry{$(3, 15)$ reg. girth 10 Alg.~\ref{alg_hqc_sc_construction_2}}
		%---------------------------------------------
		\addplot[
		color=KITgreen,
		thick,
		dashed,
		mark=square,
		mark size=3pt,
		mark options={solid}
		%forget plot,
		]
		coordinates{
			(2, 30)
			(3, 19)
			(4, 14)
			(5, 12)
			(6, 10)
		};
		\addlegendentry{$(4, 20)$ reg. girth 8 Alg.~\ref{alg_hqc_sc_construction_2}}
		%---------------------------------------------
		\addplot[
		color=KITgreen,
		thick,
		dashed,
		mark=o,
		mark size=3pt,
		mark options={solid}
		%forget plot,
		]
		coordinates{
			(4, 461)
			(5, 365)
			(6, 329)
			(7, 293)
			(8, 233)
			(9, 209)
			(10, 194)
			(11, 176)
			(12, 164)
			(13, 152)
			(14, 146)
			(15, 137)
			(16, 125)
			(17, 113)
			(19, 101)
		};
		\addlegendentry{$(4, 20)$ reg. girth 10 Alg.~\ref{alg_hqc_sc_construction_2}}
	\end{axis}
	%	\spy [blue, size=2.5cm] on (axis cs:2.8, 1e-3) in node[fill=white] at (axis cs:2.4, 1e-4);
\end{tikzpicture}

%% file: figures/comparison_dv3_dv6_g6_g10.tex
\begin{tikzpicture}
\begin{semilogyaxis}[
    width=\columnwidth,
    height=0.7\columnwidth,
    xlabel={$ {E_{\text{b}}}/{N_0} $ (dB)},    
    ylabel={BER},
     xmin=0, xmax=4,
     ymin=5e-8, ymax=2e-1,
    ymajorgrids=true,
    xmajorgrids=true,
    legend cell align={left},
    legend style={at={(1, 1)},anchor=north east,font=\scriptsize},
]
\addplot[
color={rgb,255:red,0; green,100; blue,170},
line width=.6,
mark=x,
mark size=2pt,
] 
coordinates{
	(100, 100)
};
\addlegendentry{girth 6};
\addplot[
color={rgb,255:red,223; green,155; blue,27},
line width=.6,
mark=x,
mark size=2pt,
] 
coordinates{
	(100, 100)
};
\addlegendentry{girth 10};
%\addplot[
%color={rgb,255:red,162; green,34; blue,35},
%line width=.6,
%mark=triangle,
%mark size=2pt,
%] 
%coordinates{
%	(100, 100)
%};
%\addlegendentry{code in~\cite{NB21}};
%---------------------------------------------
\addplot[
color={rgb,255:red,0; green,100; blue,170},
line width=.6,
mark=x,
mark size=2pt,
forget plot,
]
coordinates{
(0.0000e+00, 1.1955e-01)
(2.0000e-01, 1.0876e-01)
(4.0000e-01, 9.4555e-02)
(6.0000e-01, 7.8492e-02)
(8.0000e-01, 6.0110e-02)
(1.0000e+00, 3.4554e-02)
(1.2000e+00, 1.9411e-02)
(1.4000e+00, 7.5492e-03)
(1.6000e+00, 2.2849e-03)
(1.8000e+00, 6.3413e-04)
(2.0000e+00, 2.0245e-04)
(2.2000e+00, 6.8053e-05)
(2.4000e+00, 1.9836e-05)
(2.6000e+00, 7.5019e-06)
(2.7000e+00, 4.6776e-06)
(2.8000e+00, 3.0187e-06)
(2.9000e+00, 1.9043e-06)
(3.0000e+00, 1.2544e-06)
(3.1000e+00, 8.3290e-07)
(3.2000e+00, 5.9966e-07)
(3.3000e+00, 3.4607e-07)
(3.4000e+00, 2.6859e-07)
(3.5000e+00, 1.6475e-07)
};
%---------------------------------------------
\addplot[
color={rgb,255:red,0; green,100; blue,170},
line width=.6,
mark=x,
mark size=2pt,
forget plot,
]
coordinates{
(0.0000e+00, 1.1965e-01)
(2.0000e-01, 1.0858e-01)
(4.0000e-01, 9.3256e-02)
(6.0000e-01, 7.8477e-02)
(8.0000e-01, 5.6990e-02)
(1.0000e+00, 3.3023e-02)
(1.2000e+00, 1.4424e-02)
(1.4000e+00, 4.8175e-03)
(1.6000e+00, 1.3165e-03)
(1.8000e+00, 2.9837e-04)
(2.0000e+00, 6.0424e-05)
(2.2000e+00, 1.3434e-05)
(2.4000e+00, 2.8781e-06)
(2.6000e+00, 9.2513e-07)
(2.7000e+00, 5.1128e-07)
(2.8000e+00, 2.9161e-07)
(2.9000e+00, 1.8389e-07)
(3.0000e+00, 1.1423e-07)
(3.1000e+00, 8.1185e-08)
};
%---------------------------------------------
\addplot[
color={rgb,255:red,0; green,100; blue,170},
line width=.6,
mark=x,
mark size=2pt,
forget plot,
]
coordinates{
(0.0000e+00, 1.1993e-01)
(2.0000e-01, 1.0760e-01)
(4.0000e-01, 9.3656e-02)
(6.0000e-01, 7.7361e-02)
(8.0000e-01, 5.6045e-02)
(1.0000e+00, 3.3982e-02)
(1.2000e+00, 1.7090e-02)
(1.4000e+00, 5.5298e-03)
(1.6000e+00, 1.5043e-03)
(1.8000e+00, 3.9545e-04)
(2.0000e+00, 1.0421e-04)
(2.2000e+00, 3.0474e-05)
(2.6000e+00, 4.6574e-06)
(2.7000e+00, 2.9560e-06)
(2.8000e+00, 2.0501e-06)
(2.9000e+00, 1.3709e-06)
(3.0000e+00, 9.7499e-07)
(3.1000e+00, 6.7958e-07)
(3.2000e+00, 4.3917e-07)
(3.3000e+00, 3.1176e-07)
(3.4000e+00, 2.0544e-07)
(3.5000e+00, 1.2247e-07)
(3.6000e+00, 9.1100e-08)
(3.7000e+00, 5.4579e-08)
(3.8000e+00, 4.2243e-08)
};
%---------------------------------------------
\addplot[
color={rgb,255:red,0; green,100; blue,170},
line width=.6,
mark=x,
mark size=2pt,
forget plot,
]
coordinates{
(0.0000e+00, 1.1921e-01)
(2.0000e-01, 1.0830e-01)
(4.0000e-01, 9.3858e-02)
(6.0000e-01, 7.8184e-02)
(8.0000e-01, 5.7515e-02)
(1.0000e+00, 3.4451e-02)
(1.2000e+00, 1.5081e-02)
(1.4000e+00, 5.6322e-03)
(1.6000e+00, 1.4252e-03)
(1.8000e+00, 3.9057e-04)
(2.0000e+00, 8.3908e-05)
(2.2000e+00, 1.9891e-05)
(2.4000e+00, 6.7367e-06)
(2.6000e+00, 2.2783e-06)
(2.7000e+00, 1.3812e-06)
(2.8000e+00, 8.8315e-07)
(2.9000e+00, 5.5394e-07)
(3.0000e+00, 3.4921e-07)
(3.1000e+00, 2.3947e-07)
(3.2000e+00, 1.7442e-07)
(3.3000e+00, 1.1709e-07)
};
%---------------------------------------------
\addplot[
color={rgb,255:red,0; green,100; blue,170},
line width=.6,
mark=x,
mark size=2pt,
forget plot,
]
coordinates{
(0.0000e+00, 1.1938e-01)
(2.0000e-01, 1.0858e-01)
(4.0000e-01, 9.2987e-02)
(6.0000e-01, 7.8044e-02)
(8.0000e-01, 5.6888e-02)
(1.0000e+00, 3.3704e-02)
(1.2000e+00, 1.4141e-02)
(1.4000e+00, 5.5958e-03)
(1.6000e+00, 1.3473e-03)
(1.8000e+00, 2.9513e-04)
(2.0000e+00, 7.6735e-05)
(2.2000e+00, 1.6277e-05)
(2.4000e+00, 4.7398e-06)
(2.6000e+00, 1.5039e-06)
(2.7000e+00, 9.5652e-07)
(2.8000e+00, 6.4600e-07)
(2.9000e+00, 4.1007e-07)
(3.0000e+00, 2.8308e-07)
(3.1000e+00, 1.8215e-07)
(3.2000e+00, 1.2554e-07)
(3.3000e+00, 8.7048e-08)
(3.4000e+00, 6.3355e-08)
};
%---------------------------------------------
\addplot[
color={rgb,255:red,0; green,100; blue,170},
line width=.6,
mark=x,
mark size=2pt,
forget plot,
]
coordinates{
(0.0000e+00, 1.1909e-01)
(2.0000e-01, 1.0818e-01)
(4.0000e-01, 9.3581e-02)
(6.0000e-01, 7.9868e-02)
(8.0000e-01, 6.0108e-02)
(1.0000e+00, 3.9484e-02)
(1.2000e+00, 2.2329e-02)
(1.4000e+00, 1.1441e-02)
(1.6000e+00, 5.2662e-03)
(1.8000e+00, 2.7567e-03)
(2.0000e+00, 1.4405e-03)
(2.2000e+00, 8.4070e-04)
(2.4000e+00, 5.2974e-04)
(2.6000e+00, 2.9080e-04)
(2.8000e+00, 1.7376e-04)
(3.0000e+00, 1.0246e-04)
(3.2000e+00, 5.1475e-05)
(3.4000e+00, 3.1313e-05)
(3.6000e+00, 1.7351e-05)
(3.8000e+00, 8.3926e-06)
(4.0000e+00, 4.2919e-06)
(4.2000e+00, 2.2289e-06)
(4.4000e+00, 1.1039e-06)
};
%---------------------------------------------
\addplot[
color={rgb,255:red,0; green,100; blue,170},
line width=.6,
mark=x,
mark size=2pt,
forget plot,
]
coordinates{
(0.0000e+00, 1.1918e-01)
(2.0000e-01, 1.0852e-01)
(4.0000e-01, 9.4389e-02)
(6.0000e-01, 7.8304e-02)
(8.0000e-01, 5.8703e-02)
(1.0000e+00, 3.5787e-02)
(1.2000e+00, 1.6345e-02)
(1.4000e+00, 5.8917e-03)
(1.6000e+00, 1.6578e-03)
(1.8000e+00, 4.9054e-04)
(2.0000e+00, 1.4706e-04)
(2.2000e+00, 6.5709e-05)
(2.4000e+00, 2.9362e-05)
(2.6000e+00, 1.5389e-05)
(2.8000e+00, 9.1294e-06)
(3.0000e+00, 5.4200e-06)
(3.2000e+00, 3.3641e-06)
(3.4000e+00, 1.9186e-06)
(3.6000e+00, 1.1862e-06)
};
%---------------------------------------------
\addplot[
color={rgb,255:red,0; green,100; blue,170},
line width=.6,
mark=x,
mark size=2pt,
forget plot,
]
coordinates{
(0.0000e+00, 1.1896e-01)
(2.0000e-01, 1.0774e-01)
(4.0000e-01, 9.4429e-02)
(6.0000e-01, 7.7204e-02)
(8.0000e-01, 5.6072e-02)
(1.0000e+00, 3.5165e-02)
(1.2000e+00, 1.5020e-02)
(1.4000e+00, 4.9309e-03)
(1.6000e+00, 1.6416e-03)
(1.8000e+00, 3.9643e-04)
(2.0000e+00, 1.1424e-04)
(2.2000e+00, 3.6859e-05)
(2.4000e+00, 1.2662e-05)
(2.6000e+00, 5.9498e-06)
(2.8000e+00, 2.8248e-06)
(3.0000e+00, 1.3587e-06)
(3.2000e+00, 6.4014e-07)
(3.4000e+00, 2.9362e-07)
};
%---------------------------------------------
\addplot[
color={rgb,255:red,223; green,155; blue,27},
line width=.6,
mark=o,
mark size=2pt,
forget plot,
]
coordinates{
(0.0000e+00, 1.1934e-01)
(2.0000e-01, 1.0792e-01)
(4.0000e-01, 9.3950e-02)
(6.0000e-01, 7.8615e-02)
(8.0000e-01, 5.7667e-02)
(1.0000e+00, 3.3529e-02)
(1.2000e+00, 1.4336e-02)
(1.4000e+00, 4.5702e-03)
(1.6000e+00, 1.1908e-03)
(1.8000e+00, 2.7328e-04)
(2.0000e+00, 6.2337e-05)
(2.2000e+00, 1.3383e-05)
(2.4000e+00, 2.9544e-06)
(2.6000e+00, 7.7660e-07)
(2.7000e+00, 4.7723e-07)
(2.8000e+00, 2.8202e-07)
(2.9000e+00, 2.2575e-07)
(3.0000e+00, 1.4455e-07)
};
%---------------------------------------------
\addplot[
color={rgb,255:red,223; green,155; blue,27},
line width=.6,
mark=o,
mark size=2pt,
forget plot,
]
coordinates{
(0.0000e+00, 1.1984e-01)
(2.0000e-01, 1.0874e-01)
(4.0000e-01, 9.4980e-02)
(6.0000e-01, 7.7659e-02)
(8.0000e-01, 5.6086e-02)
(1.0000e+00, 3.3939e-02)
(1.2000e+00, 1.5331e-02)
(1.4000e+00, 5.0598e-03)
(1.6000e+00, 1.0964e-03)
(1.8000e+00, 2.6556e-04)
(2.0000e+00, 4.7506e-05)
(2.2000e+00, 1.0556e-05)
(2.4000e+00, 2.1585e-06)
(2.6000e+00, 5.8550e-07)
(2.7000e+00, 3.6311e-07)
(2.8000e+00, 2.3133e-07)
(2.9000e+00, 1.6172e-07)
(3.0000e+00, 1.0532e-07)
};
%---------------------------------------------
\addplot[
color={rgb,255:red,223; green,155; blue,27},
line width=.6,
mark=o,
mark size=2pt,
forget plot,
]
coordinates{
(0.0000e+00, 1.2005e-01)
(2.0000e-01, 1.0792e-01)
(4.0000e-01, 9.4469e-02)
(6.0000e-01, 7.8294e-02)
(8.0000e-01, 5.6723e-02)
(1.0000e+00, 3.3568e-02)
(1.2000e+00, 1.4402e-02)
(1.4000e+00, 4.6798e-03)
(1.6000e+00, 1.2798e-03)
(1.8000e+00, 2.8744e-04)
(2.0000e+00, 6.3340e-05)
(2.6000e+00, 7.9134e-07)
(2.7000e+00, 4.9739e-07)
(2.8000e+00, 3.0135e-07)
(2.9000e+00, 1.9228e-07)
(3.0000e+00, 1.2839e-07)
};
%---------------------------------------------
\addplot[
color={rgb,255:red,223; green,155; blue,27},
line width=.6,
mark=o,
mark size=2pt,
forget plot,
]
coordinates{
(0.0000e+00, 1.1963e-01)
(2.0000e-01, 1.0897e-01)
(4.0000e-01, 9.4740e-02)
(6.0000e-01, 7.9222e-02)
(8.0000e-01, 5.9477e-02)
(1.0000e+00, 3.4524e-02)
(1.2000e+00, 1.6968e-02)
(1.4000e+00, 5.6896e-03)
(1.6000e+00, 1.6590e-03)
(1.8000e+00, 3.9138e-04)
(2.0000e+00, 1.0212e-04)
(2.2000e+00, 2.6178e-05)
(2.4000e+00, 6.9837e-06)
(2.6000e+00, 2.0684e-06)
(2.7000e+00, 1.1432e-06)
(2.8000e+00, 6.7639e-07)
(2.9000e+00, 3.8950e-07)
(3.0000e+00, 2.3813e-07)
(3.1000e+00, 1.5294e-07)
(3.2000e+00, 1.0380e-07)
};
%---------------------------------------------
\addplot[
color={rgb,255:red,223; green,155; blue,27},
line width=.6,
mark=o,
mark size=2pt,
forget plot,
]
coordinates{
(0.0000e+00, 1.1872e-01)
(2.0000e-01, 1.0800e-01)
(4.0000e-01, 9.4756e-02)
(6.0000e-01, 7.7312e-02)
(8.0000e-01, 5.5943e-02)
(1.0000e+00, 3.5892e-02)
(1.2000e+00, 1.4955e-02)
(1.4000e+00, 4.3269e-03)
(1.6000e+00, 1.2820e-03)
(1.8000e+00, 2.4030e-04)
(2.0000e+00, 5.0637e-05)
(2.2000e+00, 8.5296e-06)
(2.4000e+00, 2.2224e-06)
(2.5000e+00, 1.0555e-06)
(2.6000e+00, 6.2517e-07)
(2.7000e+00, 3.6821e-07)
(2.8000e+00, 2.0859e-07)
(2.9000e+00, 1.6532e-07)
(3.0000e+00, 1.0817e-07)
(3.1000e+00, 7.2705e-08)
};
%---------------------------------------------
\addplot[
color={rgb,255:red,223; green,155; blue,27},
line width=.6,
mark=o,
mark size=2pt,
forget plot,
]
coordinates{
(0.0000e+00, 1.2007e-01)
(2.0000e-01, 1.0817e-01)
(4.0000e-01, 9.4347e-02)
(6.0000e-01, 7.8833e-02)
(8.0000e-01, 5.8665e-02)
(1.0000e+00, 3.4674e-02)
(1.2000e+00, 1.4877e-02)
(1.4000e+00, 4.6767e-03)
(1.6000e+00, 1.2604e-03)
(1.8000e+00, 2.9409e-04)
(2.0000e+00, 5.7166e-05)
(2.2000e+00, 1.2004e-05)
(2.4000e+00, 2.4641e-06)
(2.5000e+00, 1.2645e-06)
(2.6000e+00, 6.9967e-07)
};
%---------------------------------------------
\addplot[
color={rgb,255:red,223; green,155; blue,27},
line width=.6,
mark=o,
mark size=2pt,
forget plot,
]
coordinates{
(0.0000e+00, 1.1915e-01)
(2.0000e-01, 1.0866e-01)
(4.0000e-01, 9.4931e-02)
(6.0000e-01, 7.9282e-02)
(8.0000e-01, 5.6283e-02)
(1.0000e+00, 3.2212e-02)
(1.2000e+00, 1.4562e-02)
(1.4000e+00, 4.5835e-03)
(1.6000e+00, 9.8248e-04)
(1.8000e+00, 2.5826e-04)
(2.0000e+00, 5.3795e-05)
(2.2000e+00, 1.0915e-05)
(2.4000e+00, 2.2261e-06)
(2.5000e+00, 1.1889e-06)
(2.6000e+00, 5.9896e-07)
(2.7000e+00, 3.7555e-07)
(2.8000e+00, 2.2819e-07)
(2.9000e+00, 1.7317e-07)
};
%---------------------------------------------
\addplot[
color={rgb,255:red,223; green,155; blue,27},
line width=.6,
mark=o,
mark size=2pt,
forget plot,
]
coordinates{
(0.0000e+00, 1.2031e-01)
(2.0000e-01, 1.0678e-01)
(4.0000e-01, 9.4077e-02)
(6.0000e-01, 7.7848e-02)
(8.0000e-01, 5.5876e-02)
(1.0000e+00, 3.3971e-02)
(1.2000e+00, 1.4833e-02)
(1.4000e+00, 4.8566e-03)
(1.6000e+00, 1.1625e-03)
(1.8000e+00, 2.5518e-04)
(2.0000e+00, 5.3208e-05)
(2.2000e+00, 9.7870e-06)
(2.4000e+00, 2.4110e-06)
(2.5000e+00, 1.2445e-06)
(2.6000e+00, 7.3805e-07)
(2.7000e+00, 4.2206e-07)
(2.8000e+00, 2.5216e-07)
(2.9000e+00, 1.7202e-07)
(3.0000e+00, 1.2441e-07)
(3.1000e+00, 7.5487e-08)
(3.2000e+00, 5.3626e-08)
};
%---------------------------------------------
%\addplot[
%color={rgb,255:red,162; green,34; blue,35},
%line width=.6,
%mark=triangle,
%mark size=2pt,
%forget plot,
%]
%coordinates{
%(0.0000e+00, 1.2508e-01)
%(2.0000e-01, 1.1269e-01)
%(4.0000e-01, 9.8799e-02)
%(6.0000e-01, 8.1836e-02)
%(8.0000e-01, 6.5549e-02)
%(1.0000e+00, 4.7475e-02)
%(1.2000e+00, 3.1982e-02)
%(1.4000e+00, 2.0281e-02)
%(1.6000e+00, 1.1019e-02)
%(1.8000e+00, 6.0587e-03)
%(2.0000e+00, 3.0126e-03)
%(2.2000e+00, 1.4255e-03)
%(2.4000e+00, 6.3587e-04)
%(2.6000e+00, 2.7541e-04)
%(2.8000e+00, 1.1291e-04)
%(3.0000e+00, 4.5215e-05)
%(3.2000e+00, 1.6507e-05)
%(3.4000e+00, 6.7594e-06)
%};
\end{semilogyaxis}
\end{tikzpicture}